\documentclass[11pt]{article}   

\usepackage{lmodern}
\usepackage{graphicx}
\usepackage{booktabs}
\usepackage{amsmath,amssymb}
\usepackage{pgf}
\usepackage{bm}
\usepackage{amsfonts}
\usepackage{mathtools}
\usepackage{caption}
\usepackage{authblk}
\usepackage{natbib}
\usepackage{subfig}
\usepackage{url}
\usepackage{multirow}
\usepackage{tabularx}
\usepackage{longtable}
\usepackage{enumitem}
\usepackage{breakurl}
\usepackage{rotating}

\newcommand{\virgolette}[1]{``#1''}

\providecommand{\keywords}[1]{\vspace*{0.5cm} \noindent \textbf{Keywords:} #1.}

\newlength{\defbaselineskip}
\newcommand{\setlinespacing}[1]%
           {\setlength{\baselineskip}{#1 \defbaselineskip}}

\newcommand{\singlespacing}{\setlength{\baselineskip}{1.5 \defbaselineskip}}

\begin{document}

\title{Time and media-use of Italian Generation Y: dimensions of leisure preferences}

\author[1]{Michela Gnaldi \thanks{michela.gnaldi@unipg.it}}
\author[2]{Simone Del Sarto \thanks {simone.delsarto@email.com}}
\affil[1]{\small Department of Political Science, University of Perugia -- Via Pascoli 20 06123 Perugia (Italy)}
\affil[2]{\small Italian National Institute for the Evaluation of the Education System (INVALSI) -- Via Ippolito Nievo 35 00153 Rome (Italy)
}

\maketitle

\begin{abstract}
Time spent in leisure is not a minor research question as it is acknowledged as a key aspect of one's quality of life. The primary aim of this article is to qualify time and Internet use of Italian Generation Y beyond media hype and assumptions. To this aim, we apply a multidimensional extension of Item Response Theory models to the Italian \virgolette{Multipurpose survey on households: aspects of daily life} to ascertain the relevant dimensions of Generation Y time-use. We show that the use of technology is neither the first nor the foremost time-use activity of Italian Generation Y, who still prefers to use its time to socialise and have fun with friends in a non media-medalled manner. 
                                                                                    
\keywords{Time and media-use; Generation Y; Italian Multipurpose Survey; Multidimensional IRT models; IRT clustering algorithm}

\end{abstract}

\singlespacing
\section{Introduction}
\label{sec:intro}
In this paper we deal with time-use of Generation Y. Time spent in leisure and the instant enjoyment of leisure activities are indissolubly related to quality of life. In a recent paramount work, \cite{stig:sen:fit:09} claim the necessity to move from a measure of economic activity to a measure of individual well-being which, other than GDP, accounts for a broad range of measures of household economic activity to evaluate quality of life, such as time spent in leisure, other than intermediate goods, security, resource depletion and other important dimensions.

Generation Y, also called Net Generation, is a generation of people known to handle new media far more comprehensively than older generations. However, unexpectedly, little empirical research exists on time and media-use of this generation and currently most of what is known is based on anecdotal evidence. This study aims at filling in this gap by looking inside the black box of leisure time contained in the Italian Multipurpose Survey on Household, which provides rich and detailed information on time-use as regards to holidays, sport activities, recreational behaviours, as well as Internet use for chatting, reading, sharing political views, banking and so on. Purposely, our objective is to qualify types, or dimensions, of time-use of Italian Generation~Y. 

In Italy, research on time-use dates back to the late 90's. Much of it is concentrated on the study of the asymmetric use of time by gender. These important studies -- see, for instance, \cite{cam:fre:sab:98} -- have had the merit to uncover the uneven distribution of time among Italian men and women at a time when such a polarised allocation of time was perceived only on an anecdotal basis. In the present work, we renew the research into Italians' time-use by exploiting time-use habits of the so-called Generation Y, who attracted much attention over more recent years. The purpose of this work is to identify, classify and characterise meaningful dimensions, or categories, of time-use of Generation Y. In so doing, we pay attention to the different ways Generation Y uses time and interacts with the new media in its available time, rather than solely to the total amount of available time and the frequency of usage of a media -- which have represented the main focuses of earlier research in this field. Besides, we adopt an inductive perspective and thus leave the data reveal associations between different styles of time-use rather than a priori identify time-use categories. 

Indeed, past studies on time-use in general and, within it, on time spent interacting with Internet and the new media, tend to concentrate on measuring the amount of time, and variation in time, spent in household activities. For instance, a bulk of the North American literature \citep{ghe:bec:75, jus:sta:85,rob:god:97, agu:hur:07} reports an average leisure time increase across the whole population in the last decades. However, the question whether contemporary societies are gaining or loosing time overall is not a cutting-edge question to ask \citep{zuz:15}, as it is already known that, overall, societies gained time in the last century.

Beyond the binary differentiation of time gains and time losses or, likewise, Internet users versus non-users, there exist important variations in how people use their time and interact with the new media \citep{liv:hel:07,zil:har:09,gam:etall:16}. Overall, determining whether qualitative differences exist among users by looking at the aggregate totals of time people have available, or at the quantity of people using Internet, is unsatisfactory. As some authors agree \citep{min2:03, bra:10,  min:mon:11}, such a unidimensional approach misses important multidimensional information about people's preferences and choices. 
Therefore, characterising the nature of time and media-use into a reduced number of dimensions according to diverse time-use behaviours -- which is the purpose of this article -- is challenging as little is known about patterns of time and media-use of Generation Y.

A drawback found in the existing body of research is the lack of a common reference ground for qualifying the variety of ways in which people use both time in general and the new media in particular, and for classifying these differences into meaningful time-use types. For instance, a common category of Internet users is made of people using it for lurking or time-killing, but is qualified differently, i.e., as socialisers, joiners, lurkers, tourists. This translates in the absence of a theoretic frame of reference to characterise and interpret dimensions of time-use emerging from our own analyses. The only exemption in this context is the work of \cite{bra:10}, who run an important meta-analysis of existing studies and proposes a unified classification of types of users' behaviours. However, his study is centered on new media-use only, not on time-use in general, which is the very interest of the present work. Therefore, we will refer to the above mentioned work to characterise the dimensions of media-use emerging from our analysis, and to our own interpretation to qualify other dimensions of time-use. Further, for such an interpretation, we will be guided by some general categories of time-use purposes (i.e., recreational versus funny/entertaining) and time-use ways (i.e., individualistic versus socialising), as described in \cite{min1:03}. 

This article is organised as follows. Section \ref{sec:multidimtimeuse} discusses the multidimensional and latent nature of time-use attitude, together with the tools to deal with them. In Section \ref{sec:Multipurpose} we describe the data considered for our purpose, that is, the Italian Multipurpose Survey on Household, while  Section \ref{sec:method} describes the statistical methodology employed for this research. Section \ref{sec:appl} presents the results and some concluding remarks are given in Section \ref{sec:concl}.
\section{The multidimensional and latent nature of time-use attitude}
\label{sec:multidimtimeuse}
People's preferences and choices as regards to the way they use their time have to be treated as a multidimensional construct. For example, people may use time for socialising, having rest, volunteering, etc. In turn, each of these dimensions composing attitude towards time-use is made of a variety of single activities, i.e., having rest may imply sleeping, listening to the radio, chatting with friends, and so on. 
Further, such a construct is latent in that there is no means to directly measure it, so that it can be merely inferred from overt behaviours, which represent the construct observable manifestation (i.e., the responses to the items of a questionnaire survey on time-use). 

A number of issues arises once the latent multidimensionality of complex phenomena such as attitude towards time-use is acknowledged. The first concerns the choice and selection of the relevant dimensions which should be accounted for when studying it. When the choice and selection of the type and numbers of dimensions is carried out by means of statistical techniques, one has a vast range of available tools for data reduction. The issue is then which is the most appropriate method to infer the dimensions (i.e., time-use types, in our context) which contribute to characterise a complex phenomenon from observed single variables. 

As known, the most widespread techniques are principal component analysis and factor analysis, see for instance \cite{hot:33,har:76,McD:85}. Both are variable reduction techniques and sometimes mistaken as the same statistical method. The first involves extracting linear composites of observed variables. The second is based on a formal model predicting observed variables from latent factors. Overall, methods of data complexity reduction concerned with the structure of variables can be explorative and confirmative. Confirmative methods need to specify in advance the number and types of dimensions. Therefore, to apply them, a specific factor structure has to be specified in advance. Such a specification may be difficult to find in practice, and this happens whenever one does not have prior information on the dimensionality structure of the data at issue. Almost all existing analyses aimed at identifying time and media-use types employ one of the two above mentioned statistical techniques. While all methodologies have both advantages and limitations, their discussion is out of the scope of this paper. 

In this contribution, we focus on a technique of variable complexity reduction developed within an extended Item Response Theory (IRT) framework, that is, a hierarchical clustering algorithm developed by \cite{Bartolucci2007}, which allows us to group variables measuring the same dimension of the latent construct at issue (i.e., time-use in our context) in the same cluster. Overall, IRT models are particularly suited for the present study context, as they assume that the associations between individuals' responses are accounted for by a latent construct (i.e., students' ability, attitude towards time-use, etc.). Traditional IRT models characterise the latent person space in terms of a single unidimensional latent dimension. This implies that all variables of a measurement instrument (i.e., the items of a students' national test, the questions of a questionnaire survey on time-use) are located on the same scale, contributing to measure a single latent construct \citep{bartolucci2015statistical}. Yet, in our present case study, the survey questionnaire is composed by subsets of questions measuring potentially different latent constructs (i.e., time-use for socialising, time-use for volunteering, and so on). 

In such later contexts, the traditional IRT assumption of only one underlying latent variable is restrictive and a multidimensional extension is opportune. To overcome the unidimensional assumption, the model of \cite{Bartolucci2007} we apply here takes into account multidimensional latent traits \citep{rec:09} and more general item parameterisations than those of Rasch-type models \citep{rasch:61}, that is, the two-parameter logistic (2-PL) model introduced by \cite{birn:68}. Specifically, as it will be further specified in the following, the hierarchical clustering algorithm employed here to study the number and type of dimensions of time-use has an explorative nature and builds a sequence of nested models, starting with estimating the most general model (i.e., a multidimensional IRT model with a different dimension for each variable) and ending with the most restrictive model (i.e., a unidimensional IRT model with only one common dimension to all variables). 
\section{The Italian \virgolette{Multipurpose survey on households: aspects of daily life}}
\label{sec:Multipurpose}
In this section we describe the data used in this paper. In particular, Section \ref{subsec:info_multipurp} provides a general description of the questionnaire considered, that is, the Italian \virgolette{Multipurpose survey on households: aspects of daily life}, while Section \ref{subsec:data} focusses on the part tied with the respondents' spare time habits. 
\subsection{The Multipurpose survey}
\label{subsec:info_multipurp}
The \virgolette{Multipurpose survey on households: aspects of daily life} is an Italian survey part of an integrated system of social surveys \citep{ISTAT:16} included in the National Statistic Program. It provides information on citizens' habits, the problems they face in their every day life, satisfaction with their economic situation, functioning of public utility services, etc. Spare time is a specific aspect investigated in this survey, other than family and social life, political and social participation, health, life style, access to services.

The \virgolette{Multipurpose survey on households: aspects of daily life} is a privileged source of data which allows us to investigate the behaviour of citizens -- children, adolescents, adults and the elderly -- throughout the day in a very detailed manner, also capturing the varying degrees of combination among daily activities \citep{ISTAT:16}. The survey is primarily an objective survey, as it aims at getting detailed information on people's activities during the hours of weekdays, Saturdays, and Sundays. However, subjective questions are also included in the survey, such as those exploring citizens' satisfaction with their economic conditions and time-use availability. 

The survey has been annually conducting since 1993 and covers aspects of the Italian population's daily life. The main content areas covered by the survey may be summarised as follows: 

\begin{itemize}
\item household and population structure (i.e., household type, number of members, presence of children, etc.);
\item dwellings and residential areas structures (i.e., number of rooms in the house, green areas in the neighbourhood, acoustic and air pollution, water supply, etc.); 
\item household mobility and commuting habits;
\item household education levels and training;
\item domestic and non-domestic work; 
\item household time-use; 
\item social participation (i.e., associationism, political and religious participation); 
\item household lifestyles (i.e., eating and drinking habits);
\item household health conditions;
\item public and private household services (i.e., hospitals, transportation, administrative offices, babysitting, caregiver services etc.).
\end{itemize}
\subsection{Household time-use data in the Multipurpose survey}
\label{subsec:data}
Since the Multipurpose questionnaire has not a specific section devoted to spare time, the data record track has been accurately inspected and only the items that refer to the respondents' spare time have been selected. Overall, 31 items are considered, whose content is detailed as follows:

\begin{enumerate}[]
\item participation in scholastic recovering courses or private lessons;
\item participation in computer science courses or private lessons;
\item participation in foreign languages courses or private lessons;
\item participation in courses or private lessons on artistic and cultural activities;
\item having a period of holidays longer than four nights in the last year;
\item continuous sport activity;
\item occasional sport activity;
\item physical activity different from sport (e.g., walking, a swimming);
\item hanging out with friends;
\item listening to the radio;
\item watching television;
\item watching VHS and/or DVD;
\item using Internet for at least one of the following activities: 
sending or receiving e-mail;
calling/videochatting;
posting messages in chats, social networks, blogs, newsgroups or on-line discussion forums and using instant messaging services;
\item using Internet for at least one of the following activities: 
reading or posting opinions about social or political problems;
online attendance to consultations or votes about social or political problems;
\item using Internet for at least one of the following activities: 
reading or downloading newspapers, news, magazines;
looking for information about goods and services;
booking doctor appointments;
using trip and accommodation services;
using banking services;
\item using Internet for at least one of the following activities:
listening to the web radio;
watching TV programs on web;
watching films or videos in streaming;
\item using Internet for at least one of the following activities:
playing or downloading games, pictures, films, music;
online gaming; 
\item using Internet for at least one of the following activities:
uploading self-created contents;
building web sites or blogs;
selling goods or services;
\item using a portable device (different from computers) for at least one of the following activities: 
sending or receiving e-mails;
reading or downloading newspapers, news, magazines;
reading or downloading books or e-books:
playing or downloading games, pictures, videos, music;
social networking;
other activities;
\item using Internet for interacting with the public administration, with at least one of the following activities:
getting information from web sites;
downloading pre-edited forms;
filling pre-edited forms;
\item going to the theatre;
\item going to the cinema;
\item going to museums, exhibits, etc.;
\item going to classical music concerts or operas, or other music concerts;
\item watching live sport shows;
\item going to dance;
\item visiting archaeological sites and/or monuments;
\item reading daily newspapers at least once a week;
\item reading non-school and/or non-professional books;
\item reading weekly magazines;
\item reading non-weekly periodicals.
\end{enumerate}

Some of these items originally request a dichotomous answer, i.e.,\virgolette{yes} or \virgolette{no}, coded with 1 and 0, respectively. Therefore, the objective of such items is to know whether the respondent has or not a particular habit. Other items are multiple-choice items, hence the required answer is about the frequency with which a certain habit occurs in respondents' spare time; for example, the item \virgolette{Do you usually listen to the radio?} has three possible answers: \virgolette{no}, \virgolette{yes, everyday} and \virgolette{yes, some days}. For the purposes of the analyses reported in this paper, the multiple-choice items are dichotomised in such a way that all the answers different from \virgolette{no} are coded with 1. As a consequence, such dichotomised items assess if a specific habit is present or not in subjects' spare time. Furthermore, the items labelled from 13 to 20 are obtained grouping the single items belonging to the same multiple single-response question. Specifically, the response to each of these questions is coded with 1 if the respondent answers \virgolette{yes} to at least one of the items of the block. Finally, some responses are missing, but we hypothesise that a missing response actually stands for a negative answer (i.e., missing values are coded with 0).
\section{Statistical methodology}
\label{sec:method}
In this section we describe the statistical methodology employed in the present paper: in Section \ref{subsec:multidim_irt} a brief introduction to the statistical model is provided, while Section \ref{subsec:algorithm} details the clustering algorithm devoted to ascertain the actual number of dimensions measured by a questionnaire.
\subsection{The multidimensional Latent Class IRT model}
\label{subsec:multidim_irt}
IRT models are typical tools broadly used to analyse questionnaire responses in educational context, in order to measure respondents' latent abilities (e.g., mathematics proficiency). When this methodology is applied to other fields, we generally talk about latent traits or latent constructs. Specifically, let us suppose that we are dealing with a questionnaire made of $J$ dichotomous items, hence the response can assume value 0 or 1. This codification can have different meanings according to the context under study: for example, in the educational field, a correct answer is generally coded with 1, while 0 is used to represent a wrong response. However, in the context at issue, the questionnaire items aim at assessing if a certain activity is present or not in individuals' spare time, by responding \virgolette{yes} or \virgolette{no} to the item. Then, the meaning of each response is slightly different: if the activity asked for by an item is present, the response assumes value equal to 1, otherwise it is equal to 0.

As outlined above, to overtake the unidimensionality limit of traditional IRT models, extensions in a multidimensional way are proposed, in which the underlying latent trait measured by the questionnaire is assumed to have a multidimensional structure made of $s$ dimensions. In particular, in this paper we use the multidimensional Latent Class (LC) IRT model, proposed by \cite{Bartolucci2007}, in which it is hypothesised that the observed sample is drawn from a population divided into $k$ latent classes (i.e., sub-populations): subjects belonging to the same latent class share very similar characteristics in terms of latent trait. 

In the model at issue, we consider the \virgolette{conditional probability of success}, denoted by $\lambda_{j|c}$, that is the probability that the response to item $j$ is equal to 1 for subjects in latent class $c$. Specifically, the so-called \virgolette{two-parameter logistic} (2-PL) parametrisation of $\lambda_{j|c}$ is considered, in order to include a parameter measuring the item discrimination power. Then, the multidimensional LC \mbox{2-PL IRT} model is based on the following equation:	
\begin{equation}
\label{eq:prob_succ}
\textnormal{logit}(\lambda_{j|c}) = \gamma_j \Bigl ( \sum_{d=1}^s \delta_{jd}\theta_{cd} - \beta_j \Bigr ), \quad j=1,\ldots,J, \quad c=1,\ldots,k,
\end{equation}
\noindent where the conditional probability of success for item $j$ is parametrised using $\gamma_j$ and $\beta_j$. The former is the discrimination parameter of item $j$, while the latter is a parameter measuring the \virgolette{difficulty} of item $j$. In the educational field, this parameter represents the overall tendency to wrongly respond to item $j$, but in a general context its meaning can be translated in a sort of global attitude to respond 0 to the item. Moreover, in equation (\ref{eq:prob_succ}) $\delta_{jd}$ denotes an indicator variable equal to 1 if item $j$ contributes to measure dimension $d$ and 0 otherwise, with $d=1,\ldots,s$. Finally, $\theta_{cd}$ is the measure of the latent trait level for subjects in latent class $c$, with respect to dimension $d$.
\subsection{The hierarchical clustering algorithm}
\label{subsec:algorithm}
A crucial issue concerns the number of dimensions $s$ measured by the questionnaire. In order to find the suitable $s$ for the data under study, \cite{Bartolucci2007} proposes to use a hierarchical clustering algorithm. This procedure has an exploratory nature and aims at grouping the items that contribute to measure the same dimension. We have to note that, if a questionnaire is composed by $J$ items, the possible values of $s$ range from 1 to $J$; hence, for each value of $s$, all the possible ways to cluster the items (or groups of items) are considered, and the best one for each $s$ is selected according to a likelihood ratio (LR) test. 

Specifically, this algorithm starts estimating the most general model, that is, a model in which each item measures a different dimension, hence $s=J$. Therefore, in this initial situation the procedure estimates just one model, i.e., a $J$-dimensional model. Successively, a model with a fewer dimension than the previous one has to be estimated, thus $s=J-1$. In such situation, two items have to be aggregated, so as the $J-1$ dimensions are composed by one cluster of two items plus $J-2$ single items. Since there exist $\binom{J}{2}$ ways to cluster two items over a set of $J$ items, the same number of models can be estimated, each one with the same number of dimensions: these models only differ with respect to the composition of the dimensions in terms of items measuring them. As stated above, the best model with $J-1$ dimensions is selected considering a LR test, and so it will make in all the next steps for different values of $s$.

Thereafter, the procedure considers a further fewer dimension, hence $s=J-2$. Then, it evaluates all the possible ways to collapse the $J-1$ previous dimensions into $J-2$. It is important to underline that these dimensions can be obtained by aggregating two single items into a new group, or by clustering one single item with a previously-created group of items.

The algorithm continues estimating models with gradually fewer number of dimensions, evaluating all the possible ways to cluster items or groups of items ascertained in the previous steps, and selecting the best one for each possible number of dimensions. Finally, it ends by estimating the model with one dimension (the unidimensional one), that is the most restrictive model, in which all the items contribute to measure a unique dimension ($s=1$). 

All the item aggregations considered by the procedure steps can be graphically represented through a dendrogram, a typical representation mainly used in cluster analysis. By this tool, it is possible to see which items (or groups of items) are clustered within each step of the algorithm; in addition, this plot also shows the deviance between the initial model (with $s=J$) and the model selected at each step. 

Finally, it is necessary to choose the suitable model -- and the corresponding number of dimensions measured by the questionnaire -- among those selected in the clustering procedure. This choice depends on the level at which the dendrogram \virgolette{is cut}. We need to underline that there is no optimal criterion leading to the best choice in this regard. We can use statistical-based criteria, which consider, for example, the LR test, the Bayesian Information Criterion (BIC) index \citep{Schwarz1978}, the Akaike Information Criterion (AIC) index \citep{akaike1973information}, etc. Otherwise, the choice can be made according to subjective criteria, or knowledge based on previous studies on the same research subject. In the present paper, we use a statistical criterion based on the BIC: in particular, the final number of dimensions $s$ is selected according to the model showing the minimum BIC value.
\section{The application}
\label{sec:appl}
In this section, we discuss the results obtained by applying the methodology described in Section \ref{sec:method} on the Italian Multipurpose Surveys on Household data, with the purpose to study and characterise time and media habits of the Italian Generation Y. Specifically, Section \ref{subsec:applproc} shows the results of the clustering algorithm, while in Section \ref{subsec:applinterpretation} an interpretation of the ascertained dimensions of Generation Y time-use is provided.
\subsection{The procedure to extract dimensions of time-use preferences}
\label{subsec:applproc}
In this application, we consider the data collected in 2012. Over this year, the questionnaire was administrated to 46,464 Italian individuals, from which the data relative to Generation Y are selected according to the age of respondents. As there is a lack of agreement in the literature on the definition of Generation Y cohort starting and ending birth years, we choose to adopt a restrictive rule of units' selection by constraining our database to people aged between 18 and 24 who, at the time of the 2012 Multipurpose Survey here analysed, were born between 1988 and 1994. This choice should limit the internal heterogeneity of the cohort in terms of employment conditions, age, gender, etc. Further, the young age of our sample units should also control for a particular problem, underlined by \cite{stig:sen:fit:09}, defined involuntary unemployment, which happens when individuals cannot work as much as they would like and, as a consequence, have more available leisure than they would like. The previous choice leads us to select 3,180 subjects. 

\begin{figure}
\centering
\includegraphics[width=\textwidth, trim={0 2.5cm 0 0}, clip]{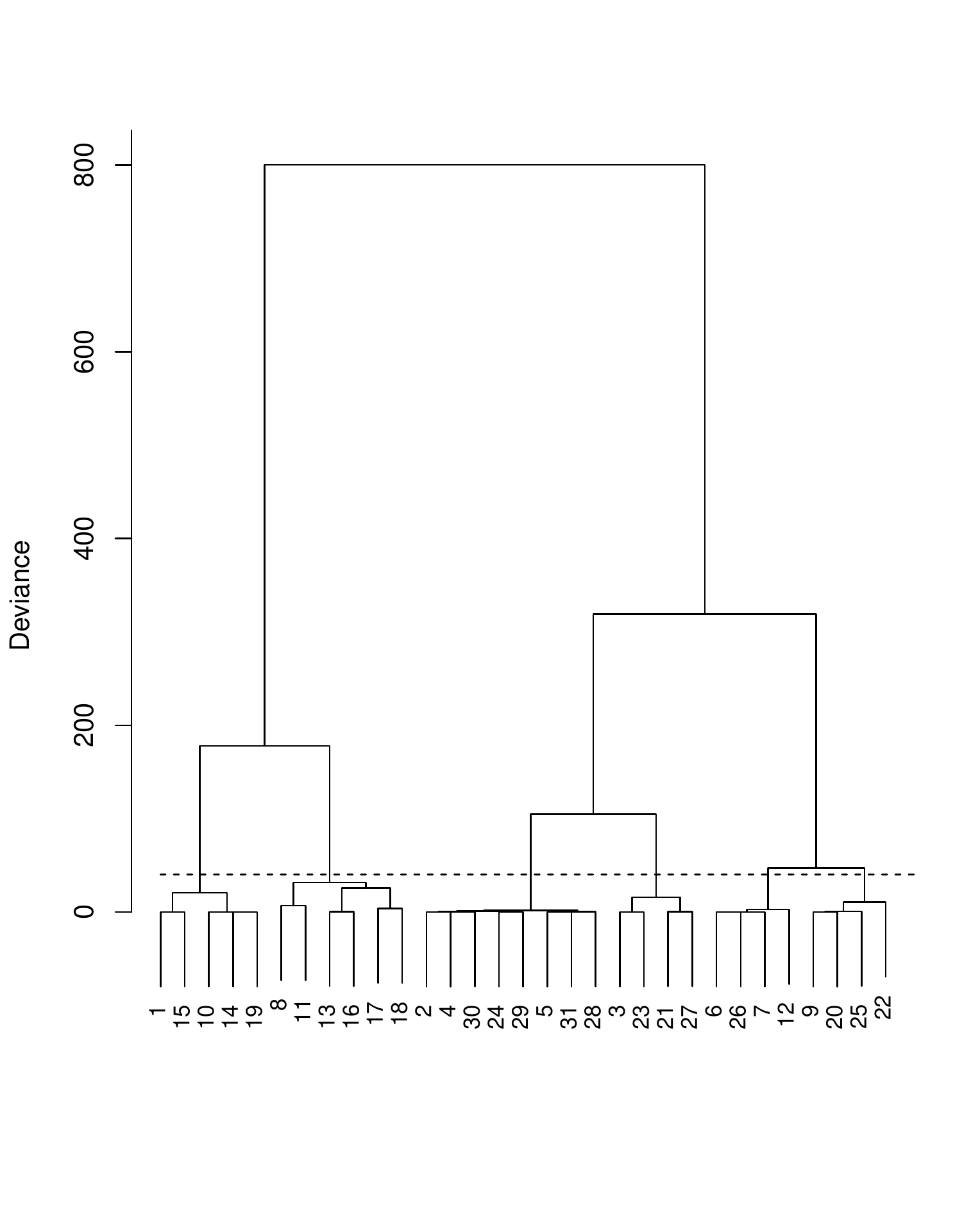}
\caption{Dendrogram resulting from the hierarchical clustering algorithm applied to a selection of the 2012 Multipurpose survey, relative to leisure habits of Generation Y. The item labels are reported on the $x$-axis, while the deviance between the model selected at each step of the clustering procedure and the initial model is reported on the $y$-axis. The level at which the dendrogram is cut is depicted with a dashed line.}
\label{fig:dend}
\end{figure}

We start with reporting the results obtained by applying the hierarchical clustering algorithm described in Section \ref{subsec:algorithm}, aimed at ascertaining the actual number of dimensions of spare time habits of Generation Y. The procedure is applied to the response pattern provided by the 3,180 units to the 31 selected items and considering $k=3$ latent classes, in accordance with previous studies investigating similar research contents \citep{gnaldietalforth}. The results are reported in Table \ref{tab:choice_s}, while the corresponding dendrogram is reported in Figure \ref{fig:dend}.

As stated above, the actual number of dimensions $s$ is selected in correspondence of the model showing the minimum BIC among all the estimated models. Therefore, this criterion leads us to cut the dendrogram at the level depicted with a dashed line in Figure \ref{fig:dend}, so as to select $s=6$ dimensions, highlighted in boldface in Table \ref{tab:choice_s}, and reported in order of appearance (from left to right) in the dendrogram. Besides, in the following we report the labels of the items that contribute to measure each of the six ascertained dimensions:  

\begin{description}
\item[Dimension 1 --] 5 items: 1, 10, 14, 15 and 19;
\item[Dimension 2 --] 6 items: 8, 11, 13, 16, 17 and 18;
\item[Dimension 3 --] 8 items: 2, 4, 5,	24,	28,	29,	30 and 31;
\item[Dimension 4 --] 4 items: 3, 21, 23 and 27;
\item[Dimension 5 --] 4 items: 6, 7, 12 and 26;
\item[Dimension 6 --] 4 items: 9, 20, 22 and 25.
\end{description}

\begin{table}
  \centering
  \caption{Output of the clustering algorithm: for each of the 30 steps, the BIC of a multidimensional 2-PL model with $s$ dimensions is reported. The final choice is reported in boldface and leads to select $s=6$ dimensions, since the corresponding model exhibits the minimum BIC among the 30 estimated models.}
    \begin{tabular}{ccc@{\qquad}ccc}
    \toprule
    step & BIC & $s$ & step & BIC & $s$ \\
    \midrule
    1     & 91,615.2 & 30    & 16    & 91,495.3 & 15 \\
    2     & 91,607.2 & 29    & 17    & 91,487.8 & 14 \\
    3     & 91,599.1 & 28    & 18    & 91,480.8 & 13 \\
    4     & 91,591.1 & 27    & 19    & 91,473.9 & 12 \\
    5     & 91,583.0 & 26    & 20    & 91,468.8 & 11 \\
    6     & 91,574.9 & 25    & 21    & 91,464.5 & 10 \\
    7     & 91,566.9 & 24    & 22    & 91,461.6 & 9 \\
    8     & 91,558.8 & 23    & 23    & 91,458.5 & 8 \\
    9     & 91,550.8 & 22    & 24    & 91,455.3 & 7 \\
    10    & 91,542.8 & 21    & 25    & \textbf{91,453.2} & \textbf{6} \\
    11    & 91,534.7 & 20    & 26    & 91,460.7 & 5 \\
    12    & 91,526.7 & 19    & 27    & 91,510.4 & 4 \\
    13    & 91,518.8 & 18    & 28    & 91,575.5 & 3 \\
    14    & 91,510.9 & 17    & 29    & 91,708.4 & 2 \\
    15    & 91,503.0 & 16    & 30    & 92,181.7 & 1 \\
    \bottomrule
    \end{tabular}
  \label{tab:choice_s}
\end{table}

\subsection{The meaning of time-use dimensions of Italian Generation Y}
\label{subsec:applinterpretation}
In this section we aim at giving a meaning to each dimension of the Italian Generation Y spare time habits, according to the content of the items (see Section \ref{subsec:data}) composing each of them. As we can see, dimension 1 and dimension 2 -- in the order of appearance (from left to right) in the dendrogram -- group the bulk of the survey questions related to Internet and portable device use. Specifically, the first dimension clusters items referring mainly to a use of Internet for information acquisition, and especially reading, and for sharing political and social views. Differently, the second dimension includes Internet activities related to posting messages in chats and social networking. Besides, watching television and other videos in streaming comes up clustered in this second dimension, while listening to the radio and participation in scholastic recovering courses and private lessons in the first. 

When referring to the general categories of time-use as described in \cite{min1:03} -- and specifically the categories of time-use purposes, that is, recreational versus fun/entertainment, and time-use ways, that is, individualistic versus socialising -- the recreational purpose seems to best characterise dimension 1, since activities grouped in this dimension (such as reading and downloading newspapers and magazines) are typically associated to a use of time for having rest and relaxing. On the other side, the fun/entertainment purpose seems to depict both the above mentioned dimensions, since activities such as playing or downloading games, pictures, videos and music are present in both dimensions. Further, while the individualistic facet seems to be the dominant motivation to drive activities (i.e., reading, attending recovering courses) clustered in dimension 1, socialisation seems instead to drive both, but especially the time-use activities (i.e., calling and videochatting) grouped in the second dimension. Accordingly, we refer to the first time-use dimension as {\em Technologically Engaging} and to the second as {\em Technologically Socialising}. 

Dimension 3 and dimension 4 both resemble traditional and, to some extend, recreational ways of spending one's time, referring to activities devoted to develop one's growth, education and culture as they group all the variables related to activities such as reading, going to concerts, museums, etc. However, dimension 3 mostly concerns activities devoted to reading and to attending recovering courses, while dimension 4 mainly refers to other active and open-space cultural activities, such as visiting museums and going to the theatre. Dimension 3, likewise dimension 1, is engaging in so that it includes time-use behaviours and activities devoted to develop one's betrothed interests and grow. However, dimension 3 misses the technological facet which is instead dominant in the first dimension. Further, while the time-use activities in dimension 3 are mostly individual activities achievable on ones' own, those clustered in dimension 4 are usually shared activities. Thus, this last dimension has a stronger socialising facet than the previous. Therefore, we denote dimension 3 as {\em Individually Engaging} and to dimension 4 as {\em Socially Engaging}. 

Dimension 5 brings together four variables referring to sportive and physical activities, going to dance and watching television. Accordingly, we shortly refer to dimension 5 as {\em Sportive}. Finally, dimension 6 collects activities which presuppose an active use of time to socialise and have fun, by sharing it with friends (i.e., going to the cinema and watching live sport shows) in a non media-medalled manner. We call this last dimension of time-use {\em Socialising Entertaining}.  

Thus, the six dimensions described above characterise Generation Y time-use habits. In order to study the occurrence of the activities included in each dimension, we compute the following quantity: 

\begin{equation}
\label{eq:lambda}
\bar{\hat{\lambda}}_d = \sum_{c=1}^k \bar{\hat{\lambda}}_{d|c} \hat{\pi}_c, \quad d=1,\ldots,s,
\end{equation}
 
\noindent where $\hat{\pi}_c$ is the estimate of the prior probability to belong to latent class $c$, while $\bar{\hat{\lambda}}_{d|c}$ is the average probability to present dimension $d$ for subjects in latent class $c$, computed taking the mean of the estimated conditional probability of success -- see equation (\ref{eq:prob_succ}) -- over the items that contribute to measure dimension $d$, with $d=1,\ldots,s$.
 
The quantity in (\ref{eq:lambda}) can be seen as the frequency with which the activities included in each dimension occur in Generation Y spare time. It is important to stress that the events \virgolette{occurrence of the activities characterising dimension $d$}, for $d=1,\ldots,s$, are not mutually exclusive, since it is possible that two or more dimensions are simultaneously present in the Generation Y time-use. As a consequence, the sum of $\bar{\hat{\lambda}}_d$ over the index $d$ can be different to 1. Therefore, we obtain the results reported in Table \ref{tab:lambda}. As we can see, the most frequent activities are those related with dimension 6 and dimension 2 ($\bar{\hat{\lambda}}_6=0.619$ and $\bar{\hat{\lambda}}_2=0.603$), which allow to state that Italian Generation Y mostly prefers to spend time in socialising, primarily in a non media-medalled manner. Furthermore, engaging in sport activities is another common occupation for Generation Y, since almost 50\% of it engages in these activities. Finally, cultural activities (dimension 4) and others devoted to develop one's growth (dimension 3) are the least frequent in Generation Y spare time.

\begin{table}
  \centering
  \caption{Description of the ascertained six dimensions of Generation Y time-use habits: for each one the labels of the items characterising them are reported, along with the frequency with which the activities included in each dimension occur, denoted with $\bar{\hat{\lambda}}_d$.}
\begin{tabular}{lclc}
\toprule
Name & $d$ & Items &  $\bar{\hat{\lambda}}_d$ \\
\midrule
{\em Socialising Entertaining} & 6 & 9, 20, 22, 25 & 0.619 \\
{\em Technologically Socialising} & 2 & 8, 11, 13, 16, 17, 18 & 0.603 \\
{\em Sportive} & 5 & 6, 7, 12, 26 & 0.488 \\
{\em Technologically Engaging} & 1 & 1, 10, 14, 15, 19 & 0.420 \\
{\em Individually Engaging} & 3 & 2, 4, 5,	24,	28,	29,	30, 31 & 0.341 \\
{\em Socially Engaging} & 4 & 3, 21, 23, 27 & 0.211 \\ 
\bottomrule
\end{tabular}
\label{tab:lambda}
\end{table}
\section{Discussion of main results}
\label{sec:concl}
The primary intent of this article is to characterise time and Internet use of Generation Y on scientific bases as, at present, most of what is known on this generation's time-use is based on media hype and assumptions. To this aim, we take advantage of the richness of information of the household time-use data contained in the Italian \virgolette{Multipurpose survey on households: aspects of daily life}. 

To analyse this dataset with the aim of grasping Generation Y main lifestyles, we make the most of extensions of IRT models to ascertain dimensions of Generation Y time-use. IRT models are particularly suitable for the objectives of this work because they assume that the associations between individuals' responses are accounted for by a latent construct, which in our case study is the unobservable attitude towards time-use of Generation Y. Specifically, the model we apply is based on a class of multidimensional IRT models. This model is an extension of basic IRT models in that it accounts for the multidimensionality of the latent construct and thus it allows us to identify meaningful categories, or dimensions, of time-use of Generation Y. Thus, in the present study, we pay attention to the different ways Generation Y interacts with time and Internet, instead of merely concentrating on the total amount of time and the frequency of usage of Internet, which have attracted much attention in previous research.

By the mentioned model, we identify six dimensions of time-use of Generation Y. We characterise such dimensions as {\em Socialising Entertaining},
{\em Technologically Socialising},
{\em Sportive},
{\em Technologically Engaging},
{\em Individually Engaging} and
{\em Socially Engaging}.
 
The first evidence emerging from our study is that the use of technology is neither the first nor the foremost time-use activity of  Italian Generation Y. Differently, Italian Generation Y devotes its time mostly to socialise and have fun, by sharing it with friends in a non media-medalled manner. The use of Internet, and the use of technology in general, come into view as an important time-use dimension but subordinately to the previously mentioned dimension, and mostly for socialisation purposes. Differently, the use of technology for information acquisition and engaging in political and social issues does not qualify much Italian Generation Y time-use. The overall picture which emerges from the present research is that socialisation -- either in a non media-medalled manner or through the use of technology -- is the distinctive trait of time-use activities of Italian Generation Y. Otherwise, engagement in time-use activities -- either in a non media-medalled manner or through the use of technology -- comes into account as the least relevant facet of time-use. 

Finally, the present study can be extended in several directions, in order to: $i.$ qualify latent classes (i.e., personality profiles) of Generation Y individuals in terms of the dimensions of time-use ascertained in the present study;  $ii.$ verify the effect of the ascertained different dimensions of time-use on the overall satisfaction with one's use of time; $iii.$ study the trend of time-use over time; $iv.$ contrast personality profiles of Generation Y, in terms of time-use, with more adult generation profiles.  
\bibliography{bibIRT}

\end{document}